\documentclass[letterpaper,12pt]{article}
\usepackage[english]{babel}
\usepackage{amssymb,amsmath}
\usepackage{endnotes}

\begin{document}

\title{What can we say about nature?}
\author{Luigi Foschini\\
\footnotesize Istituto Nazionale di Astrofisica (INAF) -- Osservatorio Astronomico di Brera\\
\footnotesize Via E. Bianchi, 46 -- 23807 -- Merate (LC) -- Italy\\
\footnotesize emails: \texttt{luigi.foschini@brera.inaf.it}; \texttt{foschini.luigi@gmail.com}
}

\date{\today}

\maketitle

\begin{abstract}
In this essay, I review the importance of languages in the study of reality, following the well-known aphorisms by Galilei, Bohr and many others. The emphasis on these aspects helps us to understand that it is not meaningful to ask if the reality if ``digital'' or ``analog'', but we have to search what is the best language to study some specific aspects of the reality. This problem is particularly felt in the case of frontier science, like quantum gravity, where, in front of several theories (syntaxes) available, there are presently neither observations nor experiments leading to the building of a convincing semantics. You will not find here recipes for a definitive theory. Just thoughts and questions. 
\end{abstract}

\scriptsize
\hangindent=6cm
\hangafter=0
\begin{flushright}
\emph{``Il ne faut pas toujours tellement \'epuiser un sujet qu'on ne laisse rien \`a faire au lecteur; il ne s'agit pas de faire lire, mais de faire penser.''}(\endnote{English translation: ``We must not always exhaust a subject, so as to leave no work at all for the reader. The matter is not to make people read, but to make them think.''}) -- 
Charles de Secondat, Baron de Montesquieu (1748)
\end{flushright}
\normalsize

\textbf{\emph{Introduction --}} It would be possible to make several examples either of an ``analog'' reality or a ``digital'' one. But the \emph{reality} is that we do not \emph{really} know what is the \emph{reality} and there is no complete and definitive answer to this question(\endnote{As you can guess, I like jokes of words.}). The most obvious obstacle to the understanding of the reality is that we all are within it, immersed in it. Nevertheless, we can speak about the nature, which is already something. As stated by Niels Bohr (cited in \cite{PETERSEN}):

\begin{quotation}
\emph{It is wrong to think that the task of physics is to find out how nature \textbf{is}. Physics concerns what can we say about nature.}
\end{quotation}

Therefore, physics -- or more generally -- science is just the writing or talking about a practice of an experience. In the XX century, the ideological or absolute visions of the world have definitely failed the comparison with quantum mechanics, general relativity and complex phenomena (chaos, self-organization and irreversibility).

The most appropriate language to speak about the nature is mathematics, as firstly noted by Galilei in the XVII century, in his famous excerpt \cite{GALILEO}:

\begin{quotation}
\emph{La filosofia \`e scritta in questo grandissimo libro che continuamente ci sta aperto innanzi a gli occhi (io dico l'Universo), ma non si pu\`o intendere se prima non s'impara a intender la lingua, e conoscer i caratteri, ne' quali \`e scritto. Egli \`e scritto in lingua matematica, e i caratteri son triangoli, cerchi, ed altre figure geometriche, senza i quali mezi \`e impossibile a intenderne umanamente parola; senza questi \`e un aggirarsi vanamente per un oscuro laberinto.}(\endnote{English translation: ``The philosophy is written in this great book that is continuously opened in front of our eyes (I say the Universe), but you cannot understand it, before knowing the language, and its characters, which was used to write it. It is written in mathematical language and the characters are triangles, circles, and other geometrical figures, and without them it is not possible to understand any word; without them it is like to round on a dark maze.''})
\end{quotation}

After Galilei, the importance of mathematics in physical sciences has been emphasized by several scientists. Just a couple of example: Rota et al. analyzed the mathematics in terms of syntax and semantics, while Bohr stressed the dependence of human beings on words. ``We are suspended in language'', he used to say (cited in \cite{PETERSEN}). From a certain point of view, mathematicians could be regarded as some kind of linguists.

However, the problem of the language of physical sciences has been often considered as a negligible detail and the above statements regarded as something like a metaphor or the result of the eccentric character of some scientist. This is likely derived from the framework of the western philosophy from Plato to Kant, which did not consider the words able to produce knowledge. Positivism, materialism, scientism were just the extremes of such kind of philosophy. Obviously, within this framework, the effectiveness of mathematics in physical sciences seems to be unreasonable \cite{WIGNER}.

The works mainly by Bohr, Heisenberg and G\"odel in the early XX century has severely challenged this ``vision of the reality'' \cite{FOSCHINI96}, although still today there is still confusion between \emph{substance} and \emph{substantive} \cite{WITTGENSTEIN} at a level that Asher Peres has felt the need of stressing that experiments occur in the laboratory and not in a Hilbert space \cite{PERES}. The title itself of the \emph{FQXi} essay contest is somehow misleading in this sense. The matter is not if the reality \emph{is} ``digital'' or ``analog'', but what is the best language -- if any -- to speak about the nature at any level of complexity. 

\textbf{\emph{Languages --}} The ability to build a language -- a system of conventional signs expressing ideas -- is natural for human beings \cite{SAUSSURE}. This ability is neither ``unreasonable'' (see \cite{WIGNER}) nor mysterious, but it depends on the unconscious elaboration of the individuals, which in turn is also the reason of the creativity \cite{HADAMARD}. It is just the ``breath'' of our minds.

The two pillars of any language are syntax and semantics. The former deals with the basic formal structure of the language, while the latter refers to the meaning of the signs and symbols of the language. Any type of language has to face with problems in understanding its structure and the meaning of the signs. This is rather obvious in disciplines as linguistics \cite{SAUSSURE} and mathematics \cite{ROTA}, but less in physics, because of the philosophical biases outlined above. Nevertheless, in physical sciences, we can identify the theory with the syntax and the experiment/observation with the semantics (e.g. \cite{FOSCHINI2}).

Generally, it is more usual to start from the semantics (observations and experiments) and then build the syntax (theory) \cite{ROTA}, although the latter is mandatory to do correct inferences and sometimes there could be the case when syntax drives the development of the semantics, like in the case of the Dirac equation and the discovery of the positron. This is important, particularly when it is necessary to study something beyond our the capacity of our experiments or observations. For example, we cannot observe what happens inside the event horizon of a spacetime singularity, but we can study it by making inferences with the theories built from the observations of the black hole effects on the nearby spacetime. 

Gian-Carlo Rota et al. noted that any syntax (axiomatic system) is just a ``window '' to see a mathematical object, which holds its identity despite the different adopted windows \cite{ROTA}. Any window can be used to emphasize one or another characteristics of the object or to find a new one, depending on the adopted perspective. When ``translated'' into physical sciences, this means that we can build different languages to speak about the same physical object, depending on which characteristic we want to study.

\textbf{\emph{One example in physics --}} Usually, this situation is exemplified by the wave/particle complementarity in quantum mechanics stated by Niels Bohr \cite{BOHR} and many others. Matter and radiation have wave- and particle-like properties, which are mutually exclusive, but complementary. However, within this context, I prefer to recall an example reported by Choudhuri in his book \cite{CHOUDHURI}, when explaining the different levels of theory used to speak about fluids. This example efficiently shows the wide choice of languages to speak about different levels of complexity in physics, spanning from quantum mechanics to fluid dynamics. I preferred this example, in order to show that the language problem is something related to any aspect of science and not only to a few research fields of frontier, like quantum mechanics or general relativity(\endnote{In addition, this example will be useful in the final part of this essay, when recalling the theory proposed by F. Markopoulou \cite{MARKOPOULOU}.}). 

If one wants to study a gas at atomic level, it is necessary to use the language of quantum mechanics. One possibility is the \emph{Schr\"odinger equation}:

\begin{equation}
\frac{ih}{2\pi}\frac{\partial \psi}{\partial t} = H\psi
\label{schroedinger}
\end{equation}

\noindent where $H$ is the quantum Hamiltonian operator of the system (generally the global energy), $h$ is the Planck constant and $\psi$ is the wave function, which in turn gives the probability density $P(x)=|\psi(x)|^2$ of finding the particle at the position $x$ (in the one-dimension case). Thus, the classical language indicating electrons around an atomic nucleus following definite orbits was no more suitable and it was necessary to speak about electrons located into ``probability clouds'' $P(x,y,z,t)$ (orbitals) calculated with the wave function $\psi(x,y,z,t)$ resulting from the solution of the Eq.~(\ref{schroedinger}). Other alternative, but rather equivalent, notations are the Heinseberg's matrices or the Dirac's bras and kets. It is worth noting that each of these alternatives emphasize certain characteristics of the quantum systems: they are windows adopted to study certain objects.

If the particles are a few and sufficiently separated to avoid quantum interferences, then it is still possible to use the Newtonian dynamics. This occurs when it is satisfied the equation:

\begin{equation}
\frac{h\sqrt[3]{n}}{\sqrt{mkT}}<<1
\label{level1}
\end{equation}

\noindent where $T$ is the temperature of the system made of $n$ particles with mass $m$. The Eq.~(\ref{level1}) is obtained by comparing the equivalent wavelenght of a particle (in term of quantum mechanics as from the complementarity wave/particle; it is the de Broglie wavelength) with the typical distances between particles, which in turn is equal to $(\sqrt[3]{n})^{-1}$. 

The equations describing the state of the $i-$th particle, identified with the canonical conjugate variables $q_i$ and $p_i$ (that are generally the position and the kinetic momentum), are the \emph{Hamilton equations}:

\begin{equation}
\label{hamilton}
\begin{split}
\frac{\partial q_i}{\partial t} &= \frac{\partial \mathcal{H}}{\partial p_i} \\
\frac{\partial p_i}{\partial t} &= -\frac{\partial \mathcal{H}}{\partial q_i}
\end{split}
\end{equation}

\noindent where $\mathcal{H}$ is the classical Hamiltonian of the system (global energy) and $i$ ranges from $1$ to a small number $n$. 

When the number $n \rightarrow \infty$, it is no more suitable (and quite meaningless) to follow the single particle dynamics and it is necessary to use the distributions or generalized functions. The density of the particles cannot be represented by a continuum function, because masses are concentrated in the particles and the empty space is between them. The only way is to use a series of ordinary functions existing only in the points where are the particles and so that their integral is constant and equal to the masses of the particles. 

In the case of a system of particles described by a distribution $f(\vec{q},\vec{p},t)$, then the evolution of $f$ is given by the \emph{Boltzmann equation}:

\begin{equation}
\frac{Df}{Dt} = \frac{\partial f}{\partial t} + \frac{\vec{p}}{m}\cdot \nabla_{\vec{q}} f  + \vec{F} \cdot\nabla_{\vec{p}} f 
\label{boltzmann}
\end{equation}

\noindent where $\vec{F}$ is an external force acting on the system and the term $Df/Dt$ is the \emph{total derivative} of $f$ (also known as \emph{convective derivative}), which in turn is just the time derivative along a trajectory in the ($\vec{q}$, $\vec{p}$) space. If the collisions among the particles are so rare to be neglected, then the term $Df/Dt=0$ and the Eq.~(\ref{boltzmann}) is named \emph{collisionless Boltzmann equation}. It is worth noting that the term ``collision'' indicate a short-range interaction between the particles, \emph{i.e.} when the distance of interaction is comparable with the size $a$ of the particles. This means that one particle can travel for a \emph{mean free path}:

\begin{equation}
\lambda_{\rm mfp} = \frac{1}{\sqrt{2}n\pi a^2}
\end{equation}

\noindent before suffering a collision. 

One of the best known solutions of the Boltzmann equation is the Maxwell distribution, obtained by considering a system not affected by external forces ($\vec{F}=0$):

\begin{equation}
f(\vec{v}) = n \left(\frac{m}{2\pi kT} \right)^{3/2} \exp \left[-\frac{m\vec{v}^2}{2kT}\right]
\label{maxwelldistribution}
\end{equation}

For a certain temperature, it is possible to calculate the most probable, the average and the root-mean square velocity of the particles:

\begin{equation}
\label{maxwellspeed}
\begin{split}
v_{\rm mp} &= \sqrt{\frac{2kT}{m}} \\
v_{\rm ave} &= \sqrt{\frac{8kT}{\pi m}} \simeq 1.13v_{\rm mp} \\
v_{\rm rms} &= \sqrt{\frac{3kT}{m}} \simeq 1.25v_{\rm mp}
\end{split}
\end{equation}

These equations are very important: they state, in the language of the kinetic theory, that the temperature is an indication of the velocity of the particles composing the system. The pressure is a kinetic effect as well, linked to the temperature by the state equation:

\begin{equation}
pV = nkT
\label{state}
\end{equation}

\noindent where $V$ is the volume of the system.

When the particle density is so great that the fluid can be approximated as a continuum, then it is necessary to adopt the hydrodynamics point of view. There is no more the single particle or a distribution of particles, but an elementary volume $\Delta V$ or an infinitesimal volume $dV$, which is always much greater than the size of atoms and molecules. The atomic or molecular structure and the interactions between particles are no more important. The infinitesimal volume $dV$ is affected by two types of forces: \emph{surface forces}, which are due to the interaction with the adjacent volumes and are described by using the stress tensor $\tau_{ij}$; \emph{mass forces}, which are due to the effects of external fields (gravitational, electric, magnetic). 

One important component of the stress tensor is the tangential component due to the viscosity, which is the cause of the friction in fluids. In the case of an ideal fluid, the stress tensor is composed only of the diagonal elements (\emph{i.e.} the force on a volume is always normal). Non-ideal fluids have also non-diagonal elements, as for example:

\begin{equation}
\tau_{xy} = -\mu \frac{dv_{x}}{dy}
\end{equation}

\noindent where $\mu$ is the viscosity coefficient. Different types of fluids are characterized by different viscosity coefficients.

To speak about a fluid, it is necessary to adopt the \emph{equation of Navier-Stokes}, which is the equivalent of the Newtonian equation for dynamics $F=ma$, but applied per unit of fluid volume:

\begin{equation}
\begin{split}
d\vec{F} & = dm\ \vec{a} \ \rightarrow \\
d\vec{F} & = \rho dV \frac{D\vec{v}}{Dt} \ \rightarrow {\rm per \ unit \ of \ volume} \\
\frac{d\vec{F}}{dV} & = \rho \frac{D\vec{v}}{Dt}
\end{split}
\label{prenavstokes1}
\end{equation}

\noindent where $\rho$ is the mass density. The term $d\vec{F}/dV$ is composed of three components per unit of volume: pressure force, sum of all the mass forces (as gradient of a scalar potential $\phi$), viscous force. By expliciting the left hand side of the Eq.~(\ref{navstokes}), it follows that:

\begin{equation}
-\frac{\nabla p}{\rho}+ \nabla \phi + \frac{\mu}{\rho}\nabla \tau_{ij} = \frac{\partial \vec{v}}{\partial t} + (\vec{v}\cdot \nabla)\vec{v}
\label{navstokes}
\end{equation}

\noindent where the term $\nu=\mu/\rho$ is also named \emph{kinematic viscosity}. 

In addition, it is necessary to take into account the \emph{continuity equation}:

\begin{equation}
\frac{\partial \rho}{\partial t} + \nabla\cdot (\rho\vec{v}) =0
\label{continuity}
\end{equation}

The Eqs.~(\ref{continuity}) and (\ref{navstokes}), together with the state equation of the fluid, such as Eq.~(\ref{state}), plus initial and boundary conditions, allow to determine the components of the velocity vector $\vec{v}$ and the pressure $p$ as a function of $(x,y,z,t)$. 

It is often used in hydrodynamics to rewrite the above equations in terms of characteristic variables: a typical length $\mathcal{L}$ and a typical velocity $\mathcal{V}$, so that the unscaled values are linked with their scaled values (indicated with a prime): 

\begin{equation}
\vec{x}=\vec{x'} \mathcal{L} \quad \vec{v}=\vec{v'} \mathcal{V}
\end{equation}

The scaled version of the Navier-Stokes equation has now two important numbers: the Reynolds number $Re=\rho\mathcal{L}\mathcal{V}/\mu$ and the Froude number $Fr=\mathcal{V}^{2}/g\mathcal{L}$, where $g$ is the gravitational acceleration (in the case the only mass force acting on the fluid is the gravitational field). The former, $Re$, is the ratio between the inertial and viscous forces, while the latter, $Fr$, is the ratio between the inertial force and the gravity. When $Re<<1$, then the viscous forces are dominating the motion of the fluid, resulting in a laminar regime. On the opposite case, $Re>>1$, then the inertial forces are dominating and the motion is quite chaotic (\emph{turbulent}). 

There are many and many more aspects that can be emphasized, depending on the adopted language, but I think that the examples shown here are more than sufficient to give an idea.

\textbf{\emph{Different windows, one reality? --}} This example explains well how it is possible to probe certain aspects of the nature by changing the language adopted to study them. It is possible to speak about water at different levels, by means of different languages offering many perspectives, but we are always speaking about water. We cannot know if what we can say about the water as it results from all these languages is an exhaustive description of this fluid, because we are part of the reality we are studying. But, at least, we can say when we have reached a complete scenario.

It is necessary to remind that the nature is incredibly complex and it is only thanks to the natural capacity of human beings to build languages that we can ride over this intricacy. However, it is not possible to use only one language, because there is no one-to-one correspondence between the syntax and the semantics. Each of them brings some treasure that has to be explicited. Although we are speaking of water, for example, we need of a proper language when we want to study certain characteristics of this fluid. 

The validity of one or another language is given by the capacity of finding regularities in the experiments or observations (i.e. physical laws, as from \cite{WIGNER}). The above equations are valid because they are linked to regularities found everywhere in space and time(\endnote{Curiously, regularities over the time are thought to allow ``predictions'', but science has wisely forgotten this bad practice as it truly became science.}): on the Earth, on Mars or any other cosmic body, in all the Universe, and \emph{as time goes by}(\endnote{Sorry, but I did not resist to remind Bogey asking Sam to play the piano in Casablanca!}). But they are offering different windows to look at the reality. Is there any of this window more fundamental than another? 

Incidentally, I note that in the XIX century, materialism and scientism have expelled philosophy from physics, but the quantum mechanics and general relativity born in the early XX century called for some metaphysics. So, philosophy was thrown out of the door, but it re-entered from the window, camouflaged as ``fundamental physics''. Paraphrasing Augustus de Morgan(\endnote{``We know that mathematicians care no more for logic than logicians for mathematics. The two eyes of exact science are mathematics and logic, the mathematical sect puts out the logical eye, the logical sect puts out the mathematical eye; each believing that it sees better with one eye than with two.'' (1868)}), I could say that today the two eyes of the knowledge are philosophy and science: scientists remove the philosophical eye, while philosophers remove the scientific eye, each believing that it sees better with one eye than with two.

\textbf{\emph{Quantum gravity --}} Fundamental physics is basically that type of frontier science, whose implications could alter significantly our philosophy of reality. Today, quantum gravity is surely the best example of such frontier physics. It should be the merging of the two fundamental physical theories, quantum mechanics and general relativity. The former deals with the ``physics of the smallest'' and is basically expressed with a ``digital'' type of language, while the latter concerns the ``physics of the fastest'' and adopts an ``analog'' language. 

Although it is not my thought, it is necessary to take also into account that perhaps there could be no possibility to merge these two languages into a comprehensive one. Perhaps, there could not be a theory more fundamental than the other, just because we need of both languages to speak completely about the reality. For example, quantum mechanics is part of the fundamental physics, but it has no meaning to say if the wave-like language is more fundamental of the particle-like one. Both languages are necessary to have the best opportunity to speak about the nature. The same could occur also below the Planck scale. So, we should not reject \emph{a priori} the possibility that quantum mechanics and general relativity could be two complementary, but mutually exclusive, languages. 

Anyway, as I stated above, I do not believe this way. I think that there could be a meaningful language best suited to speak about the reality at the level of the Planck scale, but we have not yet built it. We have not yet had the possibility to build the semantics of the Planck scale reality, which is propaedeutical to the building of a syntax. What we can do is to perform inferences based on the available languages, built with semantics referring to other aspects of the reality, and trying to design useful experiments and observations.

On the side of the semantics, the main problem is that we have presently no way to observe some conclusive effects of quantum gravity (e.g. \cite{AMELINO}, \cite{STECKER}), despite of the recent bounds set by the observations of Gamma-Ray Bursts (GRBs) performed by the \emph{Fermi} satellite that allowed to discard at least some theories (\cite{FERMI}, \cite{GHIRLANDA}, but see also \cite{GHISELLINI}, who stressed the present uncertainty in determining \emph{when} photons are generated, which affects all these measurements).  Therefore, even though there are several syntaxes proposed (see, e.g., \cite{ROVELLI} for a review), it is not possible to determine the validity of anyone. 

In addition, the syntax side itself displays the significant problem of the time. This very concept has been seldom and superficially studied up to date, within different research fields (see, e.g., \cite{ZEH} for a review). Many physical theories do not even take into account the arrow of time and are equally valid under time reversal, which is clearly unphysical. Basically, we could say that we are still at the stage of Saint Augustine, who wrote: 

\begin{quotation}
\emph{What then is time? If no one asks me, I know what it is. If I wish to explain it to him who asks, I do not know.}
\end{quotation}

Now we are forced to study in detail this concept and we cannot bypass it anymore.
Roughly speaking, time is one dimension in the four dimensional spacetime of general relativity, while it is a parameter in quantum mechanics. When trying to merge the two theories, these concepts of time clash each other. This generated two obvious different approaches: those who removed the time, which are rooted in the DeWitt-Wheeler equation (\cite{DEWITT}; e.g. loop quantum gravity, \cite{ROVELLI}), and those who removed space (e.g. \cite{MARKOPOULOU}). The latter has some aspects worth noting, in my opinion. Markopoulou proposed that the space is a property emergent above the Planck scale (``geometrogenesis''), while below it there is only time and some special kind of matter (``$\mu$atter''). This emergence is similar to the changes in properties observed when going from quantum mechanics to kinetic theory to fluid dynamics, as in the example outlined at the beginning of this essay \cite{MARKOPOULOU}. In the Markopoulou's analogy, quantum mechanics and general relativity are at a level of description similar to fluid dynamics (``low-energy theory'') and the quantum gravity theory we are searching for is at level of quantum molecular dynamics (``high-energy theory'') \cite{MARKOPOULOU}. This theory assumes that the Planck scale is not the quantum of spacetime, but it is a critical dimension when the effects described by quantum mechanics and general relativity have comparable intensities(\endnote{With reference to the example outlined in the first part of this essay, the Planck scale would have a somehow similar ``threshold'' meaning to that of Eq.~(\ref{level1}).}). 
 
On the other hand, if the Planck scale is the quantum of spacetime\endnote{This is not so straightforward, but strongly suggested by several considerations (e.g., \cite{WHEELER}, \cite{GARAY}).}, then it has no sense to study its structure, because there is not. Does a photon, the electromagnetic quantum, has a structure? Rather, we study its energy-time fluctuations. The same could be for quantum gravity, where we can study the energy-time fluctuations of the Planck scale (see \cite{WHEELER}), but we cannot study its very structure. No ``geometrogenesis'' as the emergence of matter from ``$\mu$atter'', no ``$\mu$atter'' at all, because if the quantum of spacetime (for the sake of simplicity, let's indicate it with $\aleph$, \emph{aleph}, the first letter of Hebraic alphabet) has a structure it is no more a quantum: this follows by definition. 

The ``geometrogenesis'' could instead be a process where the geometry is created by the energy-time fluctuation of $\aleph$. However, now the problem is: if a photon $\gamma$ can fluctuate with a particle and an antiparticle (e.g. $\gamma \leftrightarrows e^{-}e^{+}$), then the energy-time fluctuations of $\aleph$ generating the spacetime imply an asymmetry, because we have not (yet?) a language to speak about an ``anti-spacetime''. Wheeler \cite{WHEELER} proposed that there are fluctuations in the topology of spacetime, with creation and annihilation of handles. 

But, if this asymmetry is valid and somehow connected with the time asymmetry, then it is still possible to recover the concept of ``geometrogenesis'' as the emergence of the spacetime (but not that of ``$\mu$atter''; I stress that a quantum should be elementary, otherwise it is not a quantum), which in turn will be more like a self-organization of a large number of $\aleph$. Perhaps, the answer to our questions might rely on the possibility to use the language of irreversible thermodynamics as a bridge to connect general relativity and quantum mechanics. 

\textbf{\emph{Conclusions:}} As I have written in the abstract of this essay, I have no recipe for a final theory. The aim of this essay is to draw the attention on the linguistic issues permeating the physical science. It is not matter of saying if the reality is ``digital'' or ``analog'', but what is the best language to speak about the nature.

This is particularly important when dealing with the building of a quantum theory of gravity: at the Planck scale, the languages of quantum mechanics and general relativity are no more useful to speak about the reality, but we are not yet in the conditions to build a new language or to modify in a proper way what is already known. This occurs because the semantics of quantum gravity is not yet built, while there are already several syntaxes proposed, which are basically extensions of languages adopted in other fields, so it is not clear how much effective they are in the new context. Critical observations or experiments are strongly needed to assess this research field.

\clearpage
\theendnotes


\clearpage
\scriptsize
\begin{thebibliography}{99}
\bibitem{FERMI} Abdo A.~A. et al., ``A limit on the variation of the speed of light arising from quantum gravity effects'', \emph{Nature}, \textbf{462}, (2009), 331.

\bibitem{AMELINO} Amelino-Camelia G. \& Smolin L., ``Prospects for constraining quantum gravity dispersion with near term observations'', (2009), \texttt{arXiv:0906.3731}

\bibitem{BOHR} Bohr N., ``The quantum postulate and the recent development of atomic theory'', \emph{Nature}, \textbf{121}, (1928), p. 580-590.

\bibitem{CHOUDHURI} Choudhuri A.~R., \emph{The Physics of Fluids and Plasmas: An Introduction for Astrophysicists} (Cambridge, Cambridge University Press, 1998).  

\bibitem{SAUSSURE} de Saussure F., \emph{Cours de linguistique g\'en\'erale} (Paris, 1922). Italian translation by T. de Mauro (Bari, Laterza, 1994).

\bibitem{DEWITT} DeWitt B.~S., ``Quantum theory of gravity. I. The canonical theory'', \emph{Physical Review}, \textbf{160}, (1967), p. 1113-1148.

\bibitem{FOSCHINI96} Foschini L., ``Is science going through a critical stage?'', \emph{AEI}, \textbf{83}, (1996), p. 455-458 [\texttt{arXiv:physics/9807009}]. 

\bibitem{FOSCHINI2} Foschini L., ``On the logic of quantum physics and the concept of time'', (2001), \texttt{arXiv:quant-ph/9804040v4}

\bibitem{GALILEO} Galilei G., \emph{Il Saggiatore} (Roma, 1623). Italian translation by L. Sosio (Milano, Feltrinelli, 1979).

\bibitem{GARAY} Garay L.~J., ``Quantum gravity and minimum length'', \emph{International Journal of Modern Physics A}, \textbf{10}, (1995), p. 145-165.

\bibitem{GHIRLANDA} Ghirlanda G., Ghisellini G., Nava L., ``The onset of the GeV afterglow of GRB 090510'', \emph{Astronomy \& Astrophysics}, \textbf{510}, (2010), p. L7 (1-4).

\bibitem{GHISELLINI} Ghisellini G., Ghirlanda G., Nava L., Celotti A., ``GeV emission from gamma-ray bursts: a radiative fireball?'', \emph{Monthly Notices of the Royal Astronomical Society}, \textbf{403}, (2010), p. 926-937.

\bibitem{HADAMARD} Hadamard J., \emph{The psychology of the invention in the mathematical field} (Princeton, 1945). Italian translation by B. Sassoli (Milano, Cortina, 1993).

\bibitem{MARKOPOULOU} Markopoulou F., ``Space does not exist, so time can'', Third Prize of the FQXi 2008 Essay Contest ``The nature of time'', \texttt{arXiv:0909.1861}

\bibitem{PERES} Peres A., \emph{Quantum mechanics: concepts and methods} (Dordrecht, Kluwer, 1995).

\bibitem{PETERSEN} Petersen A., ``The philosophy of Niels Bohr'', \emph{The Bulletin of the Atomic Scientists}, September 1963, p. 8-14.

\bibitem{ROTA} Rota G.-C., Sharp D.~H., Sokolowski R., ``Syntax, semantics, and the problem of the identity of mathematical objects'', \emph{Philosophy of Science}, \textbf{55}, (1988), p. 376-386.

\bibitem{ROVELLI} Rovelli C., \emph{Quantum Gravity} (Cambridge, Cambridge University Press, 2007).

\bibitem{STECKER} Stecker F., ``Gamma-ray and Cosmic-ray tests of Lorentz invariance violation and quantum gravity models and their implications'', \emph{Science with the New Generation of High Energy Gamma-ray Experiments}, Assisi (Italy), 7-9 October 2009, edited by C. Cecchi, S. Ciprini, P. Lubrano \& G. Tosti, \emph{AIP Conf. Proc.} \textbf{1223}, (2010), p. 192-206. 

\bibitem{WHEELER} Wheeler J.~A., ``Geometrodynamics and the issue of the final state''. In: \emph{Relativity, Groups and Topology}, Les Houches Summer School of Theoretical Physics 1963, Edited by C. DeWitt and B. S. DeWitt (New York, Gordon \& Breach, 1964), p. 316-520.

\bibitem{WIGNER} Wigner E.~P., ``The unreasonable effectiveness of mathematics in the natural sciences''. In: \emph{The Collected Works of Eugene Paul Wigner -- Part B: Historical, Philosophical and Socio-Political Papers}, Vol. VI Philosophical Reflections and Syntheses, Annotated by G.~G. Emch, Edited by J. Mehra (Berlin, Springer-Verlag, 1995), p. 534-549.

\bibitem{WITTGENSTEIN} Wittgenstein L., \emph{The Blue and Brown Books} (London, 1964). Italian translation by A.~G. Conte (Torino, Einaudi, 1983). 

\bibitem{ZEH} Zeh H.~D., \emph{The physical basis of the direction of time} (Berlin, Springer-Verlag, 1989).

\end{thebibliography}
\end{document}